\documentclass[twocolumn,twoside,slac_two]{revtex4}
\usepackage{graphicx}
\usepackage{epsfig}

\usepackage{fancyhdr}
\begin{document}

\title{Search for $\gamma$-ray sources in the LAT sky with the Minimal Spanning Tree algorithm}

\author{Massaro,E.}
\affiliation{Department of Physics, University of Rome ``La Sapienza'',I-00185 Roma, ITALY}

\author{Tinebra,F.}
\affiliation{Department of Physics, University of Rome ``La Sapienza'', I-00185 Roma, ITALY}

\author{Campana,R.}
\affiliation{IASF-Roma, INAF, I-00133 Roma, ITALY} 

\author{Tosti,G.}
\affiliation{Department of Physics,University of Perugia and INFN Perugia,I-06123 Perugia, ITALY} 
\author{ on behalf of the LAT Collaboration}

\begin{abstract}
We present the topometric MST method to search for clusters of photons in
the LAT sky, which was used to obtain the seed list for the compilation 
of the First LAT catalog.
This method works well in non-dense field and can be profitably used at 
energies higher than a few GeV.
We describe the particular techniques developed by us to improve the cluster 
selection criteria and the estimate of the astronomical coordinates of the 
possibly associated gamma-ray sources.
A simulation technique to evaluate the confidence level of the source detection
is presented.

\end{abstract}

\maketitle

\thispagestyle{fancy}

\section{$\gamma$-ray sources detection with the Minimal Spanning Tree}

The \emph{Minimal Spanning Tree} (MST) is a {\it topometric} algorithm useful for searching 
clusters in a field of points (or {\it nodes}).
It was used by our group for detecting clusters of photons in $\gamma$-ray images to obtain 
a list of candidate sources for the 11-month LAT catalog.
The application of MST to LAT images, however, is not straightforward and presents some 
problems, two of which are discussed in this contribution.
The first problem concerns the accuracy of the source coordinates derived from MST, relevant 
for a safe identification of possible counterparts of $\gamma$-ray sources.
The second problem is that of estimating the significance of detection of the ``selected'' 
source candidates.
We first present a short review of MST: for a more complete description see Campana et al. (2008)\cite{campana}.

Given a set of $N$ points, or {\it nodes} ia a multidimensional space, we can compute 
the set $\{\lambda_i\}$ of weighted \emph{edges} 
connecting them: the MST (Zahn, 1971)\cite{zahn} is the tree (i.e. a graph without closed 
loops) connecting all the nodes with the minimum total weight $min[\Sigma_i \lambda_i]$.
For a set of points in a Cartesian frame, the edges are the lines joining the nodes, weighted 
by their length.

We divided the LAT sky in a few regions to take into account the presence of the Galactic 
emission and considered the photon arrival directions as nodes in a bi-dimensional graph, 
the edge weight being the angular distance between them.
Then we computed for each region the MST by means of the Prim algorithm, 
that grows a tree connecting in succession the nearest neighbour of each node already 
in the tree, 



\noindent
starting from one in the photon list.
Fig. 1 shows a field of the LAT sky, at energies higher than 4 GeV, with the MST that 
connects the events' positions.

To extract \emph{only} the locations where the photons clusterize, the following 
operations can be performed:
$i)$ {\it separation:} remove all the edges having a length $\lambda > \Lambda_{\mathrm{cut}}$, 
the separation value, defined in units of the mean edge length 
$\Lambda_{\mathrm{m}} = (\Sigma_i \lambda_i)/N$ in the MST; we obtain thus a set of 
disconnected sub-trees; 
$ii)$ {\it elimination:} remove all the sub-trees having a number of nodes 
$N \le N_{\mathrm{cut}}$: so we remove small casual clusters of photons, leaving only 
the clusters having a size over the expected flux limit.
After the application of these filters, the remaining set sub-trees $\{S_k\}$ provides 
a first list of candidate $\gamma$-ray sources.
\begin{figure*}[t]
\begin{center}
\epsfig{file=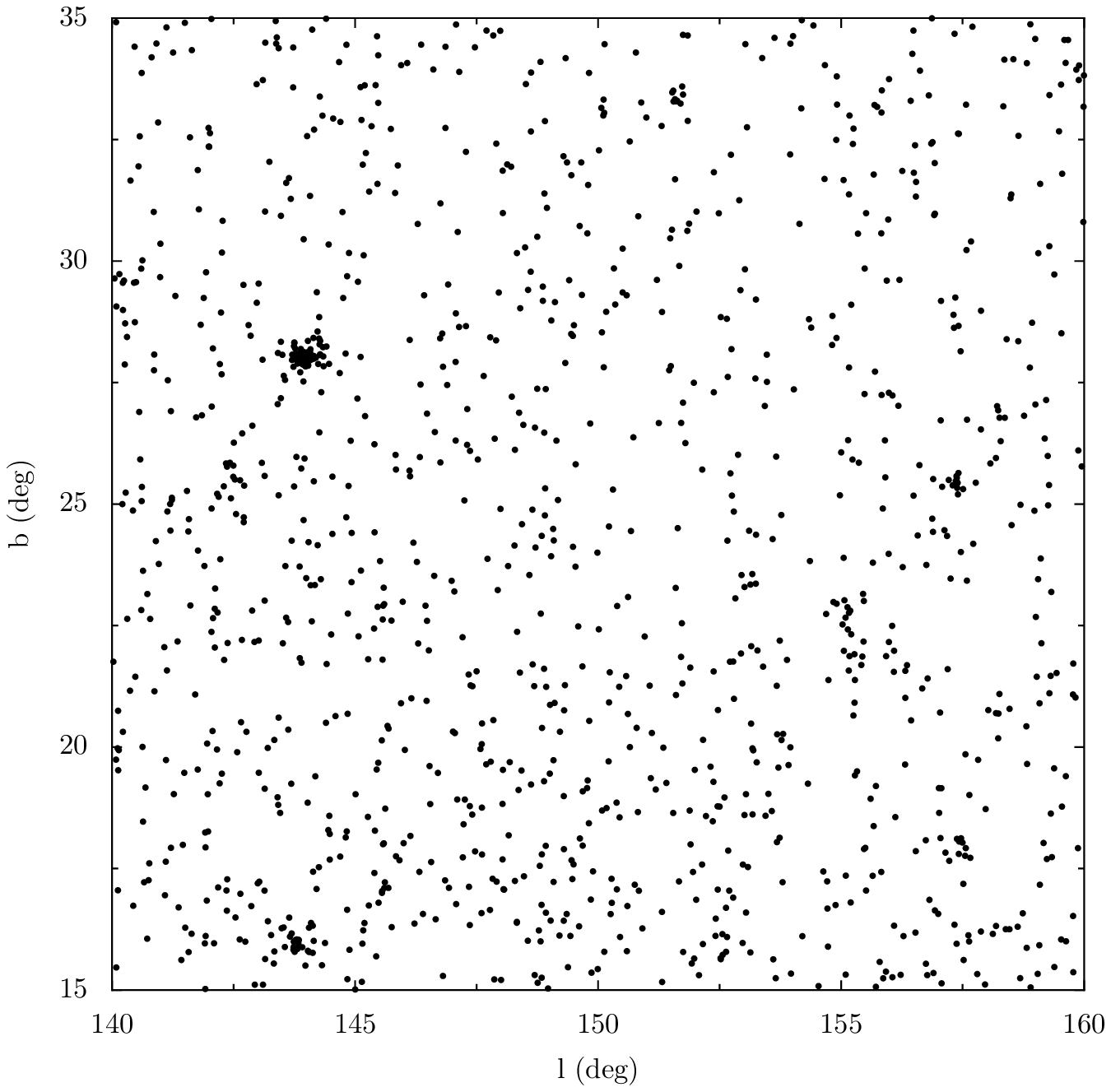,height=3.1in}\epsfig{file=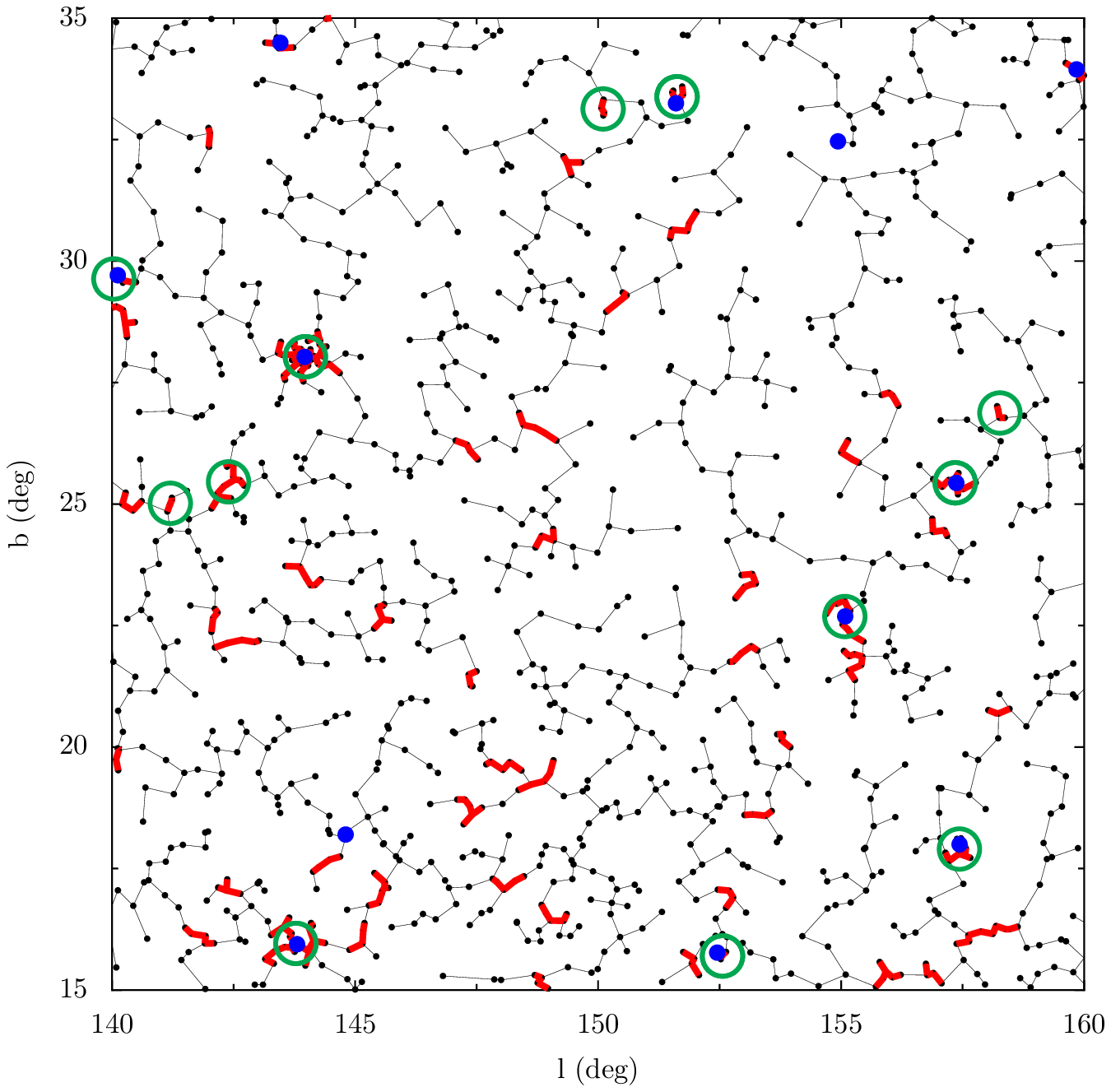,height=3.1in}
\caption{Left panel: A region (20$^{\circ} \times$ 20$^{\circ}$) of the 11 
month LAT sky at energies $>$4 GeV centered at $l$=150$^{\circ}$, $b$=25$^{\circ}$. 
Right panel: the MST between the photons and the clusters (in red) found applying the 
selection criteria $\Lambda_{cut} = 0.8 \Lambda_m$ and $N_{cut} > 3$. 
Green circles mark the candidate $\gamma$-ray sources after the selection on 
$g$ and $M$; blue dots mark the positions of sources in the LAT catalog.}
\label{fig:skymaps}
\end{center}
\end{figure*}

Campana et al. (2008)\cite{campana} considered two quantities useful to evaluate the 
``goodness'' of the clusters selected by MST: 
the number of nodes in the cluster $n_k$ and the {\it clustering parameter} $g_k$, 
defined as the ratio between $\Lambda_{\mathrm{m}}$ and $\lambda_{m,k}$, the mean 
length of the $k$-th cluster edges.
The practical application suggested the use, instead of $n_k$, of the quantity defined as
$M_k =  n_k g_k$, that we named {\it magnitude}, that combines the effects of the number 
of nodes with their clustering.

The MST is a 1-dimensional structure embedded in a 2 (or more)-dimensional space 
and therefore it can produce poor and not symmetric clusters around the centroid 
and elongated along the local tree direction.
This pattern does not fit the instrumental PSF and these clusters cannot be 
accepted as genuine candidates of $\gamma$-ray sources.
Optimized criteria to rule out these clusters have to be used.

\section{The accuracy of sources' coordinates}

A good estimate of the sources' positions in the sky is very important for the 
identification of possible counterparts.
The simple aritmetic mean of the coordinates of all photons belonging to a selected 
cluster can fail to provide a satisfactory location for different reasons: 
(\emph{i}) the cluster is in a relatively high background region and it extends in a 
particular direction with some nodes well outside the PSF; (\emph{ii}) connection 
and/or proximity with another cluster; (\emph{iii}) small number of nodes; 
(\emph{iv}) high number of nodes but moderate clustering. 

The estimate of the cluster centroids can be improved by using suitable ``weights'' 
in averaging the coordinates of the nodes or to select the nodes to be used.
We used some different approaches and verified the one producing the smallest 
differences with respect to a sample of 150 sources in the 11 month LAT catalog 
positions, chosen with different number of photons and clustering.
The adopted weights for the photon coordinates were:
$i)$ the inverse of a power of the distance to the nearest photon;
$ii)$ elimination of some of the most distant photons from the first centroid.

The mean angular difference between the LAT catalog positions (used as reference) 
and those of the MST clusters in the considered sample was 0$^{\circ}$.111.
The application of the inverse square distance weight gave reduced the mean 
difference to 0$^{\circ}$.081 while the latter of the above methods gave a 
slightly greater value around 0$^{\circ}$.095.
The chosen solution, also convenient for computational time, was that of 
weighting the distances with power of $1/\lambda$ equal to 2.

\section{The significance of source detections}
Campana et al. (2008)\cite{campana} studied the probability distributions of 
$g$ and $n$ in uniform random fields generated by a Montecarlo extraction of nodes.
A straightforward application of these results to the real sky, however, does not 
give good estimates of the probability $p_k$ of a chance detection, because of the 
presence of sources and of the Galactic emission, responsible of a not uniform 
$\gamma$-ray background over large spatial scales even at high energies.

To have a more performing significance estimate, we adopted the strategy of 
\emph{localized} Montecarlo extractions.
The wide field over which MST is computed (say, $120^{\circ}\times40^{\circ}$ 
in Galactic coordinates), was entirely divided in many small regions 
($4^{\circ}\times4^{\circ}$) and in each of them we extracted a number of nodes 
equal to that of the observed events but having a uniform random position inside 
the small region. 
In this way we retain ~the ~initial density distribution over ~the original field.
~ The MST was then

\begin{figure*}[t]
\begin{center}
\epsfig{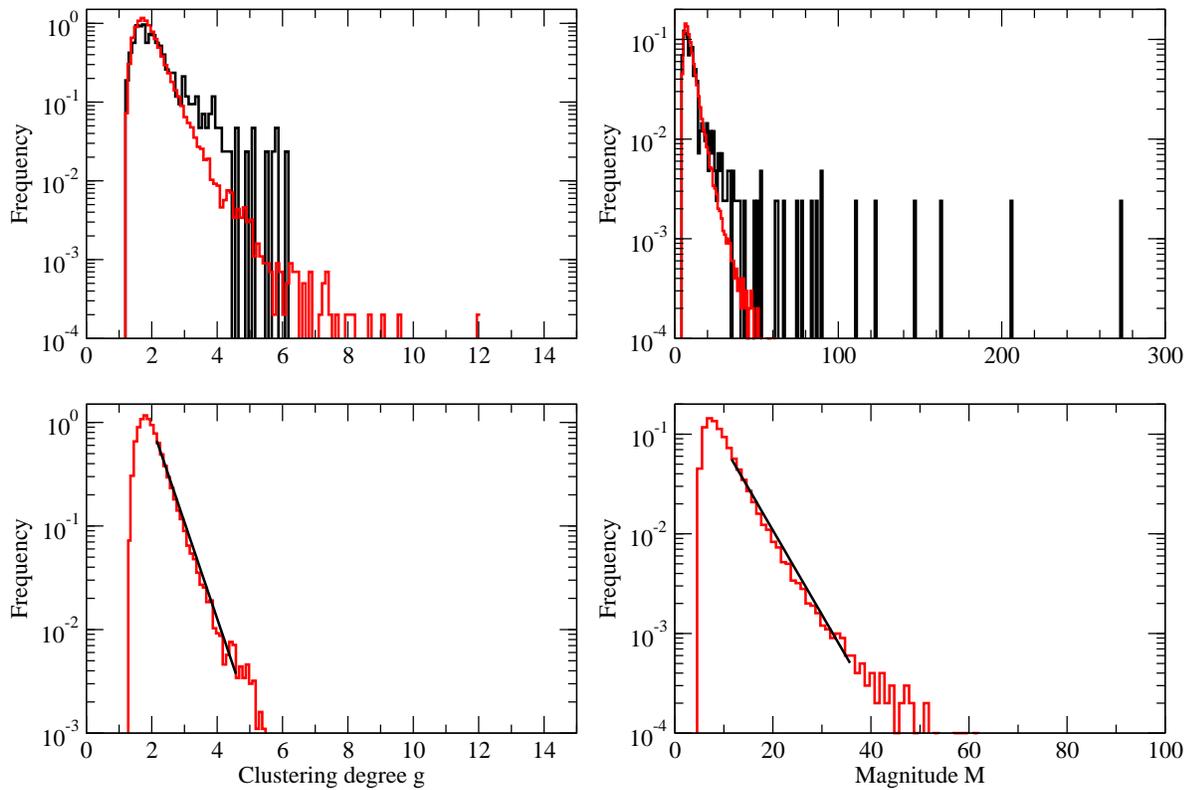}
\caption{Upper panels: Histograms of the distributions of $g$ (left) and $M$ (right) 
in the LAT field (black) and in the {\it local} Monte Carlo simulation (red).
Lower panels: Exponential best fits of the $g$ and $M$ simulated distributions (thick black lines) 
in the high value intervals.}
\label{fig:probdis}
\end{center}
\end{figure*}

\noindent
computed for the entire field and clusters were selected with the 
same filters; then we derived the distributions of $g$ and $M$. 
This procedure was repeated many time to improve the histograms of these 
distributions.
An example, where the distributions in the real field are compared with those obtained 
from the \emph{local} Montecarlo simulations. is given in Fig. 2.
Note in the real field histograms the ``long tails'' at high values of $g$ and $M$ 
corresponding to the $\gamma$-ray sources.
The significance of a source can be estimated by the distributions of simulated fields.
The portions of these curves for values higher than a percentile well above 
the mode (e.g. 60\%) do not show a large dependence on the density differences and 
are well represented by exponential laws, as shown in the lower panels of Fig. 2.
Applying a cut at the 95\% percentile to both $g$ and $M$ distributions (using the logic 
boolean OR) we made a strong selection of the clusters in Fig. 1 (right panel): practically 
all small elongated clusters, the majority of them having low $g$ or $M$ values, were rejected.
Only 12 over 55 clusters are selected as possible sources and 8 of them coincide with 
sources in the LAT catalog. 
There are few LAT sources not found by MST, but they can have soft spectra and the photon 
number above the 4 GeV threshold is too small to produce a significant cluster. 
Finally, there are four selected clusters not confirmed as detected sources in the LAT catalog.

\end{document}